\documentclass[12pt,aps,showpacs,preprintnumbers,superscriptaddress,nofootinbib]{revtex4}
\usepackage{graphicx}

\begin{document}


\newcommand{\sst}[1]{{\scriptscriptstyle #1}}
\newcommand{\beq}{\begin{equation}}
\newcommand{\eeq}{\end{equation}}
\newcommand{\beqa}{\begin{eqnarray}}
\newcommand{\eeqa}{\end{eqnarray}}
\newcommand{\dida}[1]{/ \!\!\! #1}
\renewcommand{\Im}{\mbox{\sl{Im}}}
\renewcommand{\Re}{\mbox{\sl{Re}}}
\def\simge{\hspace*{0.2em}\raisebox{0.5ex}{$>$}
     \hspace{-0.8em}\raisebox{-0.3em}{$\sim$}\hspace*{0.2em}}
\def\simle{\hspace*{0.2em}\raisebox{0.5ex}{$<$}
     \hspace{-0.8em}\raisebox{-0.3em}{$\sim$}\hspace*{0.2em}}
\def\dn{{d_n}}
\def\de{{d_e}}
\def\datom{{d_{\sst{A}}}}
\def\grhobar{{{\bar g}_\rho}}
\def\gpibar{{{\bar g}_\pi^{(I) \prime}}}
\def\gpibarz{{{\bar g}_\pi^{(0) \prime}}}
\def\gpibaro{{{\bar g}_\pi^{(1) \prime}}}
\def\gpibart{{{\bar g}_\pi^{(2) \prime}}}
\def\mn{{m_{\sst{N}}}}
\def\mx{{M_X}}
\def\mrho{{m_\rho}}
\def\qpv{{Q_{\sst{W}}}}
\def\lamtv{{\Lambda_{\sst{TVPC}}}}
\def\lamtvs{{\Lambda_{\sst{TVPC}}^2}}
\def\lamtvc{{\Lambda_{\sst{TVPC}}^3}}

\def\bra#1{{\langle#1\vert}}
\def\ket#1{{\vert#1\rangle}}
\def\coeff#1#2{{\scriptstyle{#1\over #2}}}
\def\undertext#1{{$\underline{\hbox{#1}}$}}
\def\hcal#1{{\hbox{\cal #1}}}
\def\sst#1{{\scriptscriptstyle #1}}
\def\eexp#1{{\hbox{e}^{#1}}}
\def\rbra#1{{\langle #1 \vert\!\vert}}
\def\rket#1{{\vert\!\vert #1\rangle}}

\def\lsim{{ <\atop\sim}}
\def\gsim{{ >\atop\sim}}
\def\nubar{{\bar\nu}}
\def\psibar{{\bar\psi}}
\def\Gmu{{G_\mu}}
\def\alr{{A_\sst{LR}}}
\def\wpv{{W^\sst{PV}}}
\def\evec{{\vec e}}
\def\notq{{\not\! q}}
\def\notl{{\not\! \ell}}
\def\notk{{\not\! k}}
\def\notp{{\not\! p}}
\def\notpp{{\not\! p'}}
\def\notder{{\not\! \partial}}
\def\notcder{{\not\!\! D}}
\def\notA{{\not\!\! A}}
\def\notv{{\not\!\! v}}
\def\Jem{{J_\mu^{em}}}
\def\Jana{{J_{\mu 5}^{anapole}}}
\def\nue{{\nu_e}}
\def\mn{{m_{\sst{N}}}}
\def\mns{{m^2_{\sst{N}}}}
\def\me{{m_e}}
\def\mes{{m^2_e}}
\def\mq{{m_q}}
\def\mqs{{m_q^2}}
\def\mw{{M_{\sst{W}}}}
\def\mz{{M_{\sst{Z}}}}
\def\mzs{{M^2_{\sst{Z}}}}
\def\ubar{{\bar u}}
\def\dbar{{\bar d}}
\def\sbar{{\bar s}}
\def\qbar{{\bar q}}
\def\sstw{{\sin^2\theta_{\sst{W}}}}
\def\gv{{g_{\sst{V}}}}
\def\ga{{g_{\sst{A}}}}
\def\pv{{\vec p}}
\def\pvs{{{\vec p}^{\>2}}}
\def\ppv{{{\vec p}^{\>\prime}}}
\def\ppvs{{{\vec p}^{\>\prime\>2}}}
\def\qv{{\vec q}}
\def\qvs{{{\vec q}^{\>2}}}
\def\xv{{\vec x}}
\def\xpv{{{\vec x}^{\>\prime}}}
\def\yv{{\vec y}}
\def\tauv{{\vec\tau}}
\def\sigv{{\vec\sigma}}

\def\sst#1{{\scriptscriptstyle #1}}
\def\gpnn{{g_{\sst{NN}\pi}}}
\def\grnn{{g_{\sst{NN}\rho}}}
\def\gnnm{{g_{\sst{NNM}}}}
\def\hnnm{{h_{\sst{NNM}}}}
\def\xivz{{\xi_\sst{V}^{(0)}}}
\def\xivt{{\xi_\sst{V}^{(3)}}}
\def\xive{{\xi_\sst{V}^{(8)}}}
\def\xiaz{{\xi_\sst{A}^{(0)}}}
\def\xiat{{\xi_\sst{A}^{(3)}}}
\def\xiae{{\xi_\sst{A}^{(8)}}}
\def\xivtez{{\xi_\sst{V}^{T=0}}}
\def\xivteo{{\xi_\sst{V}^{T=1}}}
\def\xiatez{{\xi_\sst{A}^{T=0}}}
\def\xiateo{{\xi_\sst{A}^{T=1}}}
\def\xiva{{\xi_\sst{V,A}}}
\def\rvz{{R_{\sst{V}}^{(0)}}}
\def\rvt{{R_{\sst{V}}^{(3)}}}
\def\rve{{R_{\sst{V}}^{(8)}}}
\def\raz{{R_{\sst{A}}^{(0)}}}
\def\rat{{R_{\sst{A}}^{(3)}}}
\def\rae{{R_{\sst{A}}^{(8)}}}
\def\rvtez{{R_{\sst{V}}^{T=0}}}
\def\rvteo{{R_{\sst{V}}^{T=1}}}
\def\ratez{{R_{\sst{A}}^{T=0}}}
\def\rateo{{R_{\sst{A}}^{T=1}}}
\def\mro{{m_\rho}}
\def\mks{{m_{\sst{K}}^2}}
\def\mpi{{m_\pi}}
\def\mpis{{m_\pi^2}}
\def\mom{{m_\omega}}
\def\mphi{{m_\phi}}
\def\Qhat{{\hat Q}}
\def\FOS{{F_1^{(s)}}}
\def\FTS{{F_2^{(s)}}}
\def\GAS{{G_{\sst{A}}^{(s)}}}
\def\GES{{G_{\sst{E}}^{(s)}}}
\def\GMS{{G_{\sst{M}}^{(s)}}}
\def\GATEZ{{G_{\sst{A}}^{\sst{T}=0}}}
\def\GATEO{{G_{\sst{A}}^{\sst{T}=1}}}
\def\mdax{{M_{\sst{A}}}}
\def\mustr{{\mu_s}}
\def\rsstr{{r^2_s}}
\def\rhostr{{\rho_s}}
\def\GEG{{G_{\sst{E}}^\gamma}}
\def\GEZ{{G_{\sst{E}}^\sst{Z}}}
\def\GMG{{G_{\sst{M}}^\gamma}}
\def\GMZ{{G_{\sst{M}}^\sst{Z}}}
\def\GEn{{G_{\sst{E}}^n}}
\def\GEp{{G_{\sst{E}}^p}}
\def\GMn{{G_{\sst{M}}^n}}
\def\GMp{{G_{\sst{M}}^p}}
\def\GAp{{G_{\sst{A}}^p}}
\def\GAn{{G_{\sst{A}}^n}}
\def\GA{{G_{\sst{A}}}}
\def\GETEZ{{G_{\sst{E}}^{\sst{T}=0}}}
\def\GETEO{{G_{\sst{E}}^{\sst{T}=1}}}
\def\GMTEZ{{G_{\sst{M}}^{\sst{T}=0}}}
\def\GMTEO{{G_{\sst{M}}^{\sst{T}=1}}}
\def\lamd{{\lambda_{\sst{D}}^\sst{V}}}
\def\lamn{{\lambda_n}}
\def\lams{{\lambda_{\sst{E}}^{(s)}}}
\def\bvz{{\beta_{\sst{V}}^0}}
\def\bvo{{\beta_{\sst{V}}^1}}
\def\Gdip{{G_{\sst{D}}^\sst{V}}}
\def\GdipA{{G_{\sst{D}}^\sst{A}}}
\def\fks{{F_{\sst{K}}^{(s)}}}
\def\FIS{{F_i^{(s)}}}
\def\fpi{{F_\pi}}
\def\fk{{F_{\sst{K}}}}
\def\RAp{{R_{\sst{A}}^p}}
\def\RAn{{R_{\sst{A}}^n}}
\def\RVp{{R_{\sst{V}}^p}}
\def\RVn{{R_{\sst{V}}^n}}
\def\rva{{R_{\sst{V,A}}}}
\def\xbb{{x_B}}
\def\mlq{{M_{\sst{LQ}}}}
\def\mlqs{{M_{\sst{LQ}}^2}}
\def\lscal{{\lambda_{\sst{S}}}}
\def\lvect{{\lambda_{\sst{V}}}}
\def\PR#1{{{\em   Phys. Rev.} {\bf #1} }}
\def\PRC#1{{{\em   Phys. Rev.} {\bf C#1} }}
\def\PRD#1{{{\em   Phys. Rev.} {\bf D#1} }}
\def\PRL#1{{{\em   Phys. Rev. Lett.} {\bf #1} }}
\def\NPA#1{{{\em   Nucl. Phys.} {\bf A#1} }}
\def\NPB#1{{{\em   Nucl. Phys.} {\bf B#1} }}
\def\AoP#1{{{\em   Ann. of Phys.} {\bf #1} }}
\def\PRp#1{{{\em   Phys. Reports} {\bf #1} }}
\def\PLB#1{{{\em   Phys. Lett.} {\bf B#1} }}
\def\ZPA#1{{{\em   Z. f\"ur Phys.} {\bf A#1} }}
\def\ZPC#1{{{\em   Z. f\"ur Phys.} {\bf C#1} }}
\def\etal{{{\em   et al.}}}
\def\delalr{{{delta\alr\over\alr}}}
\def\pbar{{\bar{p}}}
\def\lamchi{{\Lambda_\chi}}
\def\qw0{{Q_{\sst{W}}^0}}
\def\qwp{{Q_{\sst{W}}^P}}
\def\qwn{{Q_{\sst{W}}^N}}
\def\qwe{{Q_{\sst{W}}^e}}
\def\qem{{Q_{\sst{EM}}}}
\def\gae{{g_{\sst{A}}^e}}
\def\gve{{g_{\sst{V}}^e}}
\def\gvf{{g_{\sst{V}}^f}}
\def\gaf{{g_{\sst{A}}^f}}
\def\gvu{{g_{\sst{V}}^u}}
\def\gau{{g_{\sst{A}}^u}}
\def\gvd{{g_{\sst{V}}^d}}
\def\gad{{g_{\sst{A}}^d}}
\def\gvftil{{\tilde g_{\sst{V}}^f}}
\def\gaftil{{\tilde g_{\sst{A}}^f}}
\def\gvetil{{\tilde g_{\sst{V}}^e}}
\def\gaetil{{\tilde g_{\sst{A}}^e}}
\def\gvqtil{{\tilde g_{\sst{V}}^e}}
\def\gaqtil{{\tilde g_{\sst{A}}^e}}
\def\gvutil{{\tilde g_{\sst{V}}^e}}
\def\gautil{{\tilde g_{\sst{A}}^e}}
\def\gvdtil{{\tilde g_{\sst{V}}^e}}
\def\gadtil{{\tilde g_{\sst{A}}^e}}
\def\delp{{\delta_P}}
\def\delzp{{\delta_{00}}}
\def\deld{{\delta_\Delta}}
\def\dele{{\delta_e}}
\def\lnew{{{\cal L}_{\sst{NEW}}}}
\def\osffp{{{\cal O}_{7a}^{ff'}}}
\def\oszg{{{\cal O}_{7c}^{Z\gamma}}}
\def\osgg{{{\cal O}_{7b}^{g\gamma}}}


\def\slash#1{#1\!\!\!{/}}
\def\beq{\begin{eqnarray}}
\def\eeq{\end{eqnarray}}
\def\bea{\begin{eqnarray*}}
\def\eea{\end{eqnarray*}}
\def\NCA{\em Nuovo~Cimento}
\def\IJMP{\em Intl.~J.~Mod.~Phys.}
\def\NP{\em Nucl.~Phys.}
\def\PLB{{\em Phys.~Lett.}~B}
\def\JETPLett{{\em JETP Lett.}}
\def\PRL{\em Phys.~Rev.~Lett.}
\def\MPL{\em Mod.~Phys.~Lett.}
\def\PRD{{\em Phys.~Rev.}~D}
\def\PR{\em Phys.~Rev.}
\def\PRP{\em Phys.~Rep.}
\def\ZPC{{\em Z.~Phys.}~C}
\def\PTP{{\em Prog.~Theor.~Phys.}}
\def\Baryon{{\rm B}}
\def\Lepton{{\rm L}}
\def\sbar{\overline}
\def\stilde{\widetilde}
\def\st{\scriptstyle}
\def\sst{\scriptscriptstyle}
\def\vac{|0\rangle}
\def\argh{{{\rm arg}}}
\def\G{\stilde G}
\def\Wmess{W_{\rm mess}}
\def\NI{\stilde N_1}
\def\antivac{\langle 0|}
\def\infinity{\infty}
\def\mco{\multicolumn}
\def\epp{\epsilon^{\prime}}
\def\psibar{\overline\psi}
\def\nmess{N_5}
\def\chibar{\overline\chi}
\def\lagr{{\cal L}}
\def\drbar{\overline{\rm DR}}
\def\msbar{\overline{\rm MS}}
\def\conj{{{\rm c.c.}}}
\def\Et{{\slashchar{E}_T}}
\def\Etot{{\slashchar{E}}}
\def\mZ{m_Z}
\def\MPlanck{M_{\rm P}}
\def\mW{m_W}
\def\cbeta{c_{\beta}}
\def\sbeta{s_{\beta}}
\def\cW{c_{W}}
\def\sW{s_{W}}
\def\deltaeps{\delta}
\def\sigmabar{\overline\sigma}
\def\epsilonbar{\overline\epsilon}
\def\vep{\varepsilon}
\def\ra{\rightarrow}
\def\half{{1\over 2}}
\def\ko{K^0}
\def\be{\beq}
\def\ee{\eeq}
\def\bea{\begin{eqnarray}}
\def\eea{\end{eqnarray}}
\def\alr{A_{\sst{LR}}}

\def\centeron#1#2{{\setbox0=\hbox{#1}\setbox1=\hbox{#2}\ifdim
\wd1>\wd0\kern.5\wd1\kern-.5\wd0\fi
\copy0\kern-.5\wd0\kern-.5\wd1\copy1\ifdim\wd0>\wd1
\kern.5\wd0\kern-.5\wd1\fi}}
\def\ltap{\;\centeron{\raise.35ex\hbox{$<$}}{\lower.65ex\hbox{$\sim$}}\;}
\def\gtap{\;\centeron{\raise.35ex\hbox{$>$}}{\lower.65ex\hbox{$\sim$}}\;}
\def\gsim{\mathrel{\gtap}}
\def\lsim{\mathrel{\ltap}}

\def\slashchar#1{\setbox0=\hbox{$#1$}           
   \dimen0=\wd0                                 
   \setbox1=\hbox{/} \dimen1=\wd1               
   \ifdim\dimen0>\dimen1                        
      \rlap{\hbox to \dimen0{\hfil/\hfil}}      
      #1                                        
   \else                                        
      \rlap{\hbox to \dimen1{\hfil$#1$\hfil}}   
      /                                         
   \fi}                                        %

\setcounter{tocdepth}{2}


\preprint{hep-ph/0307185}

\title{Generalized Analysis of Weakly-Interacting Massive
Particle Searches}

\author{Andriy Kurylov}
\affiliation{
Division of Physics, Mathematics, and Astronomy, California
Institute of Technology, Pasadena, CA 91125}

\author{Marc Kamionkowski}
\affiliation{
Division of Physics, Mathematics, and Astronomy, California
Institute of Technology, Pasadena, CA 91125}


\begin{abstract}

We perform a generalized analysis of data from WIMP
search experiments for point-like WIMPs of arbitrary spin and
general Lorenz-invariant WIMP-nucleus interaction. We show that
in the non-relativistic limit only spin-independent (SI) and
spin-dependent (SD) WIMP-nucleon interactions survive, which can
be parameterized by only five independent parameters. We explore
this five-dimensional
parameter space to determine whether the annual modulation
observed in the DAMA experiment can be consistent with all other
experiments.  The pure SI interaction
is ruled out except for very small region of parameter space with
the WIMP mass close to 50 GeV and the ratio of the WIMP-neutron to
WIMP-proton SI couplings $-0.77\le f_n/f_p\le -0.75$. For the
predominantly SD interaction, we find an upper limit to the WIMP
mass of about 18 GeV, which can only be weakened if the constraint
stemming from null searches for energetic neutrinos from WIMP
annihilation the Sun is evaded. None of the regions of the
parameter space that can reconcile all WIMP search results can be
easily accommodated in the minimal supersymmetric extension of
the standard model.

\end{abstract}


\maketitle

\section{Introduction}
\label{sec:intro}

A weakly-interacting massive particle (WIMP) is perhaps the most
natural candidate for the dark matter that permeates our
Galactic halo \cite{kam-rev,bergstrom}.  Simply stated, if new physics at
the electroweak scale introduces a neutral stable particle, then
that particle has a cosmological density comparable to that
contributed by halo dark matter.  The most widely
studied possibility for the WIMP is the neutralino, a linear
combination of the supersymmetric partners of the photon and
$Z^0$ and Higgs boson, the lightest superpartner (LSP) in many
minimal supersymmetric extensions of the standard model (MSSMs).
However, other possibilities include the sneutrino (the
neutrino's superpartner), heavy neutrinos, particles in
non-minimal supersymmetric models, and Kaluza-Klein modes in
models with universal extra dimensions; there may be others as
well.

If such WIMPs exist in the Galactic halo, they may be detected
directly via detection of the $O(30\,{\rm keV})$ recoil energy imparted to
a nucleus in a low-background detector when a halo WIMP elastically
collides with the nucleus \cite{witten85,wasserman}.  They may
also be detected indirectly
via observation of energetic neutrinos (i.e., $E_\nu\gtrsim10$
GeV) produced by annihilation of WIMPs that have accumulated in
the Sun and/or Earth \cite{SOS}.  A third possibility is observation of
cosmic-ray antiprotons, positrons, or gamma rays produced by
annihilation of dark-matter particles in the Galactic halo,
although the certainty with which we can predict the fluxes of
such cosmic rays is limited by considerable uncertainties in the
distribution of dark matter in the halo as well as in cosmic-ray
propagation.

Of the several direct dark-matter-detection experiments underway,
one---the DAMA collaboration---has reported an annual modulation
in their NaI detector that they attribute to the difference in the
WIMP flux incident on the Earth between the summer and winter
\cite{dama-naI}.  Since their announcement, several other
experiments \cite{cdms,edel,zeplin-1,dama-Xe} have reported null
results that they argue conflict with the DAMA detection.  In
order to come to this conclusion, however, it is assumed that the
WIMP has only scalar (spin-independent; SI) interactions with
nucleons, and with equal interaction strength with neutrons and
protons.  Although such interactions are favored in most of the
supersymmetric models that appear in the literature, they are by
no means universal to the MSSM, nor to WIMP models more generally.
Ref. \cite{kam-vogel} studied the possibility that the DAMA
results might be due to a WIMP with an axial-vector
(spin-dependent; SD) coupling to either protons or to neutrons,
and showed that the former possibility conflicts with null
searches for energetic neutrinos from the Sun, while the latter is
inconsistent with other direct-detection experiments.  However,
although the realm of possibilities that have been explored has
been expanded, it is still not universal.

In this paper we expand further on prior analyses of
WIMP-detection experiments and explore whether the DAMA modulation
may be consistent with null searches in a more generalized
parameter space.  We suppose that the WIMP is a pointlike
particle, but with arbitrary spin.  We consider particles that are
either self-charge-conjugate, or that have an antiparticle, either
with or without a particle-antiparticle asymmetry. We begin in the
next Section by showing that from the point of view of dark-matter
detection, the Dirac structure of the interaction Lagrangian of
all such particles is such that the interactions relevant for
dark-matter detection can be parameterized by only five
quantities: the particle mass, SI couplings to neutrons and
protons, and SD couplings to protons and neutrons.  In Section III
we review the cross sections.  Section IV discusses the
experimental results we use from the DAMA NaI experiment, the DAMA
Xe experiment, EDELWEISS, ZEPLIN-I, and energetic-neutrino
searches \cite{SK1}.\footnote{We do not explicitly consider the
constraints from the CDMS experiment \cite{cdms} as they are
surpassed by the constraints from the EDELWEISS experiment
\cite{edel} in the region of interest; both CDMS and EDELWEISS
used cryogenic Ge detectors.} In Section V we search the
five-dimensional parameter space for regions that are consistent
with all of the experimental results. The results are summarized
in Section VI, the Conclusions.  The bottom line is that there are
regions in this five-dimensional parameter space for both SD and
SI interactions that are consistent with all current experimental
results, but these regions are unlikely to occur in any MSSM and
do not apply to any of the other WIMP candidates that have
appeared in the recent literature.

\section{Generalized WIMP-Nucleon Interaction}
\label{sec:theory}

WIMP searches have usually been interpreted within the framework
of the minimal supersymmetric standard model (MSSM) because it
contains an excellent WIMP candidate, the lightest neutralino.
Interactions of the neutralino with quarks (and ultimately with
nucleons and nuclei) have been extensively studied in the
literature (see {\it e.g.} \cite{kam-rev,vogel92}). For
slow-moving Majorana neutralinos, the neutralino-nucleon
interaction has only two terms,
\beq
\label{eq:chi-lagr}
{\cal L}_\chi&=&
{\bar \chi}\gamma^\mu\gamma_5\chi {\cal J}_\mu^5(x)
+{\bar \chi} \chi {\cal S}(x)~,\nonumber \\
{\cal J}_\mu^5(x)&=& \sqrt{2}G_F\left(a_p {\bar p}\gamma_\mu \gamma_5 p
+a_n {\bar n}\gamma_\mu \gamma_5 n \right )~,\nonumber \\
{\cal S}(x)&=& f_p {\bar p} p + f_n {\bar n} n~,
\eeq
where $G_F$ is the Fermi constant, and $a_{p,n}$ and $f_{p,n}$
are, respectively, the proton and neutron coupling constants given
in Ref.~\cite{kam-rev}.\footnote{ We absorbed the coupling
constants into the definitions of ${\cal J}_\mu^5$ and ${\cal S}$
to make this simple notation consistent with the convention
adopted in Ref. \cite{kam-rev}.} The term containing ${\cal S}(x)$
gives rise to the spin-independent (SI) interaction, whereas the
${\cal J}_\mu^5(x)$ term is responsible for the spin-dependent
(SD) interaction.

Although Eq.~(\ref{eq:chi-lagr}) was derived assuming that WIMPs
are slow-moving Majorana spin-$1/2$ fermions within the MSSM, it
can be generalized to particles of arbitrary spin, independent of
whether they are self-charge-conjugate or not. Consider a general
Lorentz-invariant interaction for the WIMP-nucleon coupling,
\beqa
\label{eq:lagr-general}
{\cal L}_\chi&=& \left({\cal S}_\chi+{\cal
P}_\chi\right)\left(G_s{\bar N}N +G_p{\bar N}\gamma_5 N\right)\nonumber \\
&+&\left({\cal V}_\chi^\mu+{\cal A}_\chi^\mu\right)
\left(G_v{\bar N}\gamma_\mu N+G_a\left\{{\bar N}\gamma_\mu\gamma_5
N+2M_N{q_\mu\over m_\pi^2-q^2}{\bar N}\gamma_5 N\right\}\right) \nonumber \\
&+&\left({\cal T}_\chi^{\mu\nu}+{\cal D}_\chi^{\mu\nu}\right)
\left(G_t{\bar N}\sigma_{\mu\nu} N +G_d{\bar N}\sigma_{\mu\nu}\gamma_5
N\right)~.
\eeqa
Here, ${\cal S}_\chi$ and ${\cal P}_\chi$ are linear combinations
of scalar and pseudoscalar operators built from the $\chi$ field,
${\cal V}_\chi^\mu$ and ${\cal A}_\chi^\mu$ are the corresponding
vector and axial-vector operators, and ${\cal T}_\chi^{\mu\nu}$
and ${\cal D}_\chi^{\mu\nu}$ are the tensor and pseudotensor
operators. In the Weyl representation, the Dirac structures appearing
in Eq.~(\ref{eq:lagr-general}) are (Latin indices indicate spatial
components),
\beqa \label{eq:Diracology}
\gamma^0&=&\left( \begin{array}{cc} 0&1\\
1&0
\end{array}\right),\qquad \gamma^i=\left( \begin{array}{cc}
0&\sigma^i\\ -\sigma^i&0 \end{array}\right),\qquad \gamma_5=\left(
\begin{array}{cc} -1&0\\ 0&1 \end{array}\right)~,\nonumber \\
\sigma^{\mu\nu}&=&{i\over
2}\{\gamma^\mu\gamma^\nu-\gamma^\nu\gamma^\mu\}~, \eeqa
where $\sigma^i$ are Pauli matrices. In order to use this
interaction to calculate matrix elements for WIMP-nucleus
scattering, one must sum over all nucleons in the target nucleus.

The WIMP-nucleon interaction is determined by the underlying
WIMP-quark interaction, and to obtain Eq.~(\ref{eq:lagr-general})
from the underlying interaction one must take its matrix elements
between one-nucleon states. In the limit $|\vec q|\ll M_N$
relevant for dark-matter detection, one can do this by simply
replacing the quark fields with the nucleon fields and rescaling
the corresponding coupling constants. For example, $g_V^q\langle
N|{\bar q}\gamma_\mu q|N\rangle=G_V^q{\bar u_N}\gamma_\mu u_N$,
where $u_N$ is the nucleon spinor. This procedure holds for all
operators except for the quark axial-vector current whose matrix
element has a pole:
\beqa
\label{eq:pcac}
\langle N|J_{q5}^\mu|N\rangle&=&{\bar
N}\left(g_A\gamma^\mu\gamma^5+g_P(q^2)q^\mu\gamma_5\right)N~,
\nonumber \\
g_P(q^2)&=&g_A{2M_N\over m_\pi^2-q^2}~,
\eeqa
where the second line implements the PCAC hypothesis, and
$m_\pi$ is the pion mass. This argument justifies the appearance of
the particular structure multiplying $G_a$ in
Eq.~(\ref{eq:lagr-general}). The second term in the first line
of Eq.~(\ref{eq:pcac}) becomes important if $|\vec q|\gsim
m_\pi$ \cite{vogel92}. For simplicity, we do not explicitly keep
the pion pole term in the following. However, it should be
understood that this term always appears in the form given by
the last term in the second line of
Eq.~(\ref{eq:lagr-general}).

Except for the pion pole term, Eq.~(\ref{eq:lagr-general}) does not contain
operators with explicit factors of momentum transfer $q$. Such terms are
considered subleading. As will be shown below, such operators are suppressed by
powers
of $q/M_{p}$ or $q/M_\chi$.
It is easy to show that in the frame where the nucleus is
initially at rest the components of $q$ are related by $2q_0
M_{nuc}+q^2=0$, which also shows that $q_0\ll|{\vec
q}|$. Therefore, we find,
\beq
\label{eq:suppress}
{q\over M_{p}}\sim \sqrt{{q_0M_{nuc}\over M_{p}^2}}=\sqrt{{q_0 A\over
M_p}}\lsim {0.01\sqrt{A}}
\lsim 0.1~,
\eeq
where $A\lsim 100$ is the atomic number of the nucleus and $q_0$ is the recoil
energy, which we assume not to significantly exceed 100 keV.

Let us look at each of the one-nucleon operators appearing in Eq.
(\ref{eq:lagr-general}) and determine which of them have
non-vanishing nuclear matrix elements in the non-relativistic limit.\\
\indent {\bf Scalar operators.} The scalar operator ${\bar N} N$
simply counts nucleons; it obviously survives in the
non-relativistic limit. On the other hand, nuclear matrix elements
of the pseudoscalar operator ${\cal P}_N=\sum_k{\bar
N_k}(x+r_k)\gamma_5 N_k(x+r_k)$ are suppressed: $\langle f|{\cal
P}_N|i \rangle\sim |\vec q|/M_{p}$. Here, $x$ and $r_k$
correspond, respectively, to the center of mass of the nucleus and
to the position of the $k$th nucleon relative to the center of
mass. Since our discussion is not affected by the fact that the
nucleus has a finite size we can consider it to be a point
particle and set all $r_k$ to zero. In a more accurate treatment
one can account for finite size $R$ by introducing a form-factor
$F(|{\vec q}|R)$. Defining an \lq\lq axial-vector current"
operator ${\hat A}_5^\mu(x)=\int^x {\cal P}_N(x)dx^\mu$ as an
intermediate step we obtain:
\beqa
\label{eq:pseudoscalar}
\langle f|{\cal P}_N(x)|i\rangle
&=&\langle f|\partial_\mu{\hat A}_5^\mu(x)| i\rangle
=iq_\mu e^{iq\cdot x}\langle f|{\hat A}_5^\mu(0)|i\rangle \nonumber \\
&\approx& iq_i e^{iq\cdot x} P\langle f|{\hat J_N^i}|i\rangle
+{\cal O}(q^2)~,
\eeqa
where ${\hat J}_N$ is the nuclear spin operator. In the last step
we used the fact that spin is the only axial vector describing a
state of the nucleus that does not vanish as $q\to 0$. The scalar
constant $P$ has the dimension of inverse mass. The largest
quantity that has this property and is not singular in the limit
$q\to 0$ is the inverse mass of the nucleon. Therefore, $P\sim
1/M_{p}$, and matrix elements of ${\cal P}_N$ are suppressed
according to Eq.~(\ref{eq:suppress}). Another way to see that
$P\sim 1/M_p$ (and, {\it e.g.}, not $1/M_{nuc}$) is to first
consider the one-nucleon operator ${\bar N}\gamma_5 N$ using free
Dirac spinors for the nucleon fields. Using free spinors is a good
approximation because the binding energy of a nucleon in a nucleus
is much smaller than the nucleon mass, which means that nucleons
are only slightly off-shell. A direct calculation yields,
\beq
{\bar N}(p+q)\gamma_5 N(p)={q_i \over M_p}\eta^{\dagger\prime}\sigma_i\eta~,
\eeq
where $\eta,\eta^\prime$ are Weyl spinors characterizing the
states of the nucleon with the initial momentum $p$, and
$\sigma_i$ is the vector of Pauli matrices. In a nucleus one
obtains
\beq {q_i\over M_p}\langle f|\sum_k
\eta^{\dagger\prime}_k\sigma_i\eta_k|i \rangle \sim {q_i\over
M_p}\langle f|{\hat J_N^i}|i\rangle~, \eeq
in agreement with Eq.~(\ref{eq:pseudoscalar}).
\\
\indent {\bf Vector operators.} For ${\cal V}_N^\mu=\sum_k{\bar
N_k}\gamma^\mu N_k$ we must have:
\beq
\label{eq:vector}
\langle f|{\cal
V}_N^\mu|i\rangle =V^+_N(p_f+p_i)^\mu+V^-_N(p_f-p_i)^\mu~,
\eeq
where $V^+_N$ and $V^-_N$ are constants. Here,
$p_i=(M_{nuc},\vec 0)$ and $p_f=q+p_i$. Just like for ${\cal
P}_N$, we may conclude on dimensional grounds that
$V^+_N,V^-_N\sim 1/M_{p}$. Then the time component of ${\cal
V}_N$ is of order unity, whereas the spatial components are
suppressed as ${\cal V}_N^i\sim q^i/M_{p}$. The time component
${\cal V}_N^0$ multiplies ${\cal V}_{\chi,0}$ and ${\cal
A}_{\chi,0}$ in Eq.~(\ref{eq:lagr-general}), which transform,
respectively, as a scalar and a pseudoscalar under the extended
rotational group. Therefore, they can be effectively absorbed into
${\cal S}_\chi$ and ${\cal P}_\chi$ in the non-relativistic case.

For the axial-vector operator ${\cal A}_{N5}^\mu=\sum_k{\bar
N_k}\gamma^\mu\gamma_5 N_k$ the situation is reversed. The time
component transforms as a pseudoscalar under the extended
rotational group and its matrix elements are suppressed for the
same reason as those of ${\cal P}_N$. The spatial components
transform like a pseudovector leading to
\beq
\langle f|{\cal A}_{N5}^i|i\rangle=A_N
\langle f|{\hat J_N^i}|i\rangle~,
\eeq
where $A_N$ is a dimensionless constant, which can in general be
of order unity.\\
\indent {\bf Tensor operators.} Let us first consider the operator
${\cal T}_N^{\mu\nu}=\sum_k{\bar N_k}\sigma^{\mu\nu} N_k$. Nonzero
components of $\sigma^{\mu\nu}$ are $\sigma^{0i}$ and
$\sigma^{ij}$, $i,j=1,2,3$. Under the extended rotational group,
$\sigma^{0i}$ transforms as a polar vector. Therefore, in analogy
to Eq.~(\ref{eq:vector}),
\beq
{\cal
T}_N^{0i}=T^+_N(p_f+p_i)^i+T^-_N(p_f-p_i)^i~,
\eeq
which is suppressed similarly to ${\cal V}_N^i$ in
Eq.~(\ref{eq:vector}). To analyze ${\cal T}_N^{ij}$ consider its
dual ${\cal T}_N^i=\epsilon^{ijk}{\cal T}_{N,jk}=\sum_k{\bar
N_k}\epsilon^{ijk}\sigma_{jk} N_k\equiv 2\sum_k{\bar N_k}\sigma^i
N_k$, where $\epsilon^{ijk}$ is the three-dimensional Levi-Civita
tensor. In the last step, we used the property
$\sigma_{jk}=\epsilon^{jkl}\sigma_l$, which can be derived from
Eq.~(\ref{eq:Diracology}). We obtain
\beq
\label{eq:nr-tensor}
\langle f|{\cal T}_N^{ij}|i \rangle={1\over 2} \epsilon^{ijk} {\cal T}_{N,j}
=\epsilon^{ijk}\langle f|\sum_k{\bar N_k}\sigma^i
N_k|i \rangle=T_N\epsilon^{ijk}\langle f|{\hat J}_{N,k}|i \rangle~,
\eeq
where $T_N$ is a dimensionless constant. Therefore, ${\cal
T}_N^{ij}$ survives in the non-relativistic limit.

The operator ${\cal D}_N^{\mu\nu}=\sum_k{\bar
N_k}\sigma^{\mu\nu}\gamma_5 N_k$ can be expressed as
$(i/2)\epsilon^{\mu\nu\rho\lambda}{\cal T}_{N,\rho\lambda}$, which
follows from the identity
$\sigma^{\mu\nu}\gamma_5 =
(i/2)\epsilon^{\mu\nu\rho\lambda}\sigma_{\rho\lambda}$.
Since only ${\cal T}_N^{ij}$ survives in the non-relativistic
limit we conclude that ${\cal D}_N^{0i}$ are the only non-vanishing
components of ${\cal D}_N^{\mu\nu}$ in this limit:
\beq
\label{eq:nr-pseudoT}
{\cal D}_N^{0i}={i\over 2} \epsilon^{0i jk} {\cal T}_{N,jk}
={i\over 4}\epsilon^{0i jk}\epsilon_{jkl} {\cal T}_N^l
={iT_N\over 2}\langle f|{\hat J}_N^i|i \rangle~.
\eeq

The antisymmetric tensor $\epsilon_{ijk}$ in
Eq.~(\ref{eq:nr-tensor}) is contracted with spatial components of
$\cal{T^{\mu\nu}_\chi}$ and $\cal{D^{\mu\nu}_\chi}$ in
Eq.~(\ref{eq:lagr-general}). The corresponding quantities ${\cal
T}^{ij}_\chi\epsilon_{ijk}$ and ${\cal D}^{ij}_\chi\epsilon_{ijk}$
transform, respectively, as an axial and a polar vector under the
extended rotational group. Therefore, in the non-relativistic
limit ${\cal T}^{ij}_\chi\epsilon_{ijk}$ can be absorbed into
${\cal A}_{\chi,k}$ and ${\cal D}^{ij}_\chi\epsilon_{ijk}$ into
${\cal V}_{\chi,k}$. Similarly, taking the Levi-Civita structure
$\epsilon^{0i jk}\epsilon_{jkl}$ from Eq.~(\ref{eq:nr-pseudoT}) we
conclude that ${\cal
T}_{\chi,0i}\epsilon^{0ijk}\epsilon_{jkl}\equiv -2 {\cal
T}_{\chi,0l}$ transforms as a polar vector and ${\cal
D}_{\chi,0i}\epsilon^{0ijk}\epsilon_{jkl}\equiv -2 {\cal
D}_{\chi,0l}$ as an axial vector. They can be absorbed into,
respectively, ${\cal V}_{\chi,l}$ and ${\cal A}_{\chi,l}$. With
the above considerations in mind Eq.~(\ref{eq:lagr-general}) can
be effectively rewritten in the non-relativistic limit as
\beqa
\label{eq:lagr-general-NR}
{\cal L}_\chi&=& \left({\cal S}_\chi+{\cal P}_\chi\right)G_s{\bar N}N
+\left({\cal V}_\chi^i+{\cal A}_\chi^i\right) G_a{\bar
N}\gamma_i\gamma_5 N~,
\eeqa
where ${\cal S}_\chi,~{\cal P}_\chi,~{\cal V}_\chi^i,~{\rm
and}~{\cal A}_\chi^i$ have been redefined to absorb the
contributions from the vector and tensor interactions. Moreover,
since the general arguments used for ${\cal P}_N$ and ${\cal
V}_N^i$ also apply to ${\cal P}_\chi$ and ${\cal V}_\chi^i$
(with
the suppression $|\vec q|/M_\chi\sim 10^{-3}$ in this case) we can
safely neglect the latter.

The scalar density ${\cal S_\chi}(x)$ in
Eq.~(\ref{eq:lagr-general-NR}) can be written as $\chi^\dagger
\chi$. For a WIMP of spin $J_\chi$, $\chi$ is a
$(2J_\chi+1)$-component non-relativistic spinor ({\it e.g.} in the
MSSM, $\chi$ is a two-component Weyl spinor). The spatial
components of the axial-vector density ${\cal A}_\chi^l(x)$ in the
non-relativistic limit are proportional to the spin density
$\chi^\dagger{\hat S}_\chi^l\chi$. Finally, we arrive at the
following form of the WIMP-nucleon interaction Lagrangian in the
non-relativistic limit,
\beqa \label{eq:lagr-short-NR} {\cal L}_\chi&=&
4f_N\chi^\dagger\chi\eta_N^{\dagger}\eta_N +16\sqrt{2}G_F a_N
\chi^\dagger{\vec S}_\chi\chi\eta_N^{\dagger}{\vec S_N} \eta_N
+{\cal O}\left({q\over M_{p,\chi}}\right)~, \eeqa
where $\eta_N$ is the two-component Weyl spinor for the nucleon
(initial and final state spinors may be different), $N=n,p$,
${\vec S_N}=\vec\sigma/2$, and the notation for the couplings has
been adjusted to match that of Eq.~(\ref{eq:chi-lagr}). In this
form, Eq.~(\ref{eq:lagr-short-NR}) is valid for non-relativistic
point-like WIMPs of arbitrary spin.

The operators neglected in Eq.~(\ref{eq:lagr-short-NR}) are
suppressed by a factor of $0.01\sqrt{A}$ relative to the leading
order terms [see Eq.~(\ref{eq:suppress}))]. Strictly speaking, in
neglecting such terms we made an implicit assumption that the
couplings $f_N$ and $G_Fa_N$ in Eq. (\ref{eq:lagr-short-NR}) are
not suppressed relative to the other couplings in
Eq.~(\ref{eq:lagr-general}), such as $G_p$. If it turned out that
$G_p\gsim 100\sqrt{A} f_N$ then one would have to consider the
${\bar N}\gamma_5 N$ operator even though it is formally
suppressed in the non-relativistic limit. In this paper we ignore
this possibility, and modulo the aforementioned assumption our
analysis is model-independent.

\subsection{Some Examples}
\label{sec:examples}

Below we consider some examples of theories that have WIMPS with
spin-0, spin-1/2 (Dirac particle), and spin-1 and explicitly show
that in each case the WIMP-nucleon interaction reduces to
Eq.~(\ref{eq:lagr-short-NR}).

\subsubsection{Spin-0}

It is possible that in a supersymmetric theory the lightest SUSY
particle is the sneutrino $\tilde \nu$, the scalar partner of
the neutrino. In the non-relativistic limit, such a particle
interacts with a nucleon as \cite{witten85},
\beqa
{\cal L}_{\tilde \nu}&=&-\sqrt{2}G_F J_{\tilde \nu}^\mu J_{N,\mu}~, \nonumber
\\
J_{\tilde \nu}^\mu&=& \left(p^{\prime \mu}_{\tilde \nu}+p^\mu_{\tilde
\nu}\right){\tilde \nu}^\dagger{\tilde \nu}~, \nonumber \\
J_N^\mu&=&\left(T_N^3-2Q_N\sin^2\theta_W\right){\bar N}\gamma^\mu N~,
\eeqa
where $p^{\mu}_{\tilde \nu},p^{\prime\mu}_{\tilde \nu}$ are the
initial and final momentum of $\tilde \nu$, $T_N^3$ and $Q_N$ are
the isospin and the electric charge of the nucleon, and $\theta_W$
is the Weinberg angle. For a slow-moving Dirac fermion of mass $M$
the leading terms in the $v/c$ expansion of the $u$- and
$v$-spinors are (in the Weyl representation),
\beq
\label{eq:nr-spinors}
u(p)=\left( \begin{array}{c} \left(1+{{\vec p}
\cdot {\vec \sigma}\over 2M_p}\right)\eta\\
\left(1-{{\vec p}\cdot{\vec \sigma}\over 2M}\right)\eta \end{array}
\right),\qquad
v(p)=\left( \begin{array}{c}
\left(1+{{\vec p}\cdot{\vec \sigma}\over 2M_p}\right)\eta\\
-\left(1-{{\vec p}\cdot{\vec \sigma}\over 2M_p}\right)\eta
\end{array} \right)~, \eeq
where $p$ is the fermion momentum and $\eta$ is a Weyl spinor.
Using the above expressions one can explicitly verify that only
the time component of the nucleon vector current survives in the
limit $\vec q\to 0$. Therefore, one is left with a purely SI
interaction, which is a special case of
Eq.~(\ref{eq:lagr-short-NR}) with
$f_N=-\sqrt{2}G_F\left(T_N^3-2Q_N\sin^2\theta\right)$ and $a_N=0$.

\subsubsection{Spin-1/2}

In  theories with extra dimensions Kaluza-Klein (KK) excitations
of the SM particles can produce viable dark matter candidates. As
an example, consider a Dirac KK neutrino, which interacts with
nucleons via $Z^0$ exchange as \cite{Appelquist02},
\beq
{\cal L}_{\nu}=-{\sqrt{2}G_F}\left(T_N^3-2Q_N\sin^2\theta_W\right){\bar
\nu}_{KK}\gamma_\mu \nu_{KK} {\bar N}\gamma^\mu N~.
\eeq
Just like in the previous example, only time components of
currents remain in the non-relativistic limit, and we again obtain
Eq.~(\ref{eq:lagr-short-NR}) with
$f_N=-\sqrt{2}G_F\left(T_N^3-2Q_N\sin^2\theta_W\right)$ and $a_N=0$.

\subsubsection{Spin-1}

Kaluza-Klein excitations of the SM gauge bosons could constitute
cold dark matter \cite{feng02}. If, for example, dark matter is
composed of excitations of the hypercharge gauge boson $B_1$ the
WIMP-quark interaction has the form:
\beqa
\label{eq:spin-1}
{\cal L}_B&=&-\left(\beta_q+\gamma_q\right)B_1^{\dagger\mu}B_{1\mu}{\bar q}q-i
\alpha_q B^\dagger_{1\mu}B_{1\nu}\epsilon^{0\mu\nu\rho}{\bar
q}\gamma_\rho\gamma_5 q~, \\
\alpha_q&=& {e^2\over 2\cos^2\theta}\left[{Y_{qL}^2m_{B_1}\over
m_{q_L^1}^2-m_{B_1}^2}
+(L\to R)\right]~,\nonumber \\
\beta_q&=&E_q{e^2\over 2\cos^2\theta}\left[Y_{qL}^2{m_{B_1}^2+m_{q_L^1}^2\over
(m_{q_L^1}^2-m_{B_1}^2)^2}
+(L\to R)\right]~,\nonumber \\
\gamma_q&=&m_q{e^2\over 4\cos^2\theta}{1\over m_h^2}~,
\eeqa
where $Y_{q}$ is the hypercharge of the quark with mass $m_q$ and
energy $E_q$, $m_{q^1}$ is the mass of the quark's first KK
excitation, and $m_h$ is the Higgs boson mass. The first term in
Eq.~(\ref{eq:spin-1}) corresponds to the SI interaction in
Eq.~(\ref{eq:lagr-short-NR}):
\beqa
\langle N|{\bar q}q|N\rangle&=&{M_N\over m_q}f_{T_q}^N{\bar N}N=2{M_N\over
m_q}f_{T_q}^N\eta^\dagger_N\eta_N~, \nonumber \\
f_N&=&{1\over 2}\sum_{u,d,s,\dots}(\beta_q+\gamma_q){M_N\over
m_q}f_{T_q}^N~, \eeqa
where $f_{T_q}^N$ for various quark flavors can be found {\it
e.g.} in Refs. \cite{kam-rev,vogel92}, and $f_N$ is normalized to
reproduce Eq.~(\ref{eq:lagr-short-NR}). The second term produces a
SD WIMP-nucleon interaction:
\beqa
i\epsilon^{0ijk}B^\dagger_{1i}B_{1j}\langle N|{\bar q}\gamma_k\gamma_5
q|N\rangle
&=&-4\Lambda_q^N\langle B_1|{\vec S}_B|B_1\rangle{\bar N}\vec S_N N~,\nonumber
\\
a_N&=&{1\over 4\sqrt{2}G_F}\sum_{u,d,s,\dots}\alpha_q\Lambda_q^N,
\eeqa
where $\left({\hat S}_B^k\right)_{ij}=i\epsilon^{0kij}$ is the matrix of the
spin operator for $B_1$, $\Lambda_q^N$ is a dimensionless constant of order
unity, and $a_N$ is normalized to reproduce Eq.~(\ref{eq:lagr-short-NR}). A
recent analysis gives $\Delta_u^p=\Delta_d^n=0.78\pm0.02$,
$\Delta_d^p=\Delta_u^n=-0.48\pm0.02$, and $\Delta_s^p=\Delta_s^n=-0.15\pm0.02$
\cite{mallot00}.

\section{WIMP-Nucleus Cross Section}
\label{sec:crossection}

The SI and SD WIMP-proton cross sections at low momentum transfers
are easily calculated from
Eq.~(\ref{eq:lagr-short-NR})\footnote{For a
non-self-charge-conjugate WIMP, such as heavy Dirac neutrino, the
WIMP-proton cross section must be multiplied by an extra factor of
1/4. This modification does not affect
Eq.~(\ref{eq:x-sections}).},
\beqa
\sigma_{\chi p}^{SI}&=&{4f_p^2\over \pi}M_{red}^2(M_p)~, \nonumber \\
\sigma_{\chi p}^{SD}&=&{128J_\chi(J_\chi+1)J_p(J_p+1)
G_F^2a_p^2\over 3\pi}M_{red}^2(M_p)~, \nonumber \\
M_{red}(M)&=&{M_\chi M\over M_\chi + M}~,
\eeqa
where $M_\chi$ and $J_\chi$ are the WIMP mass and spin. In our
convention, the non-relativistic spinors $\psi$ for both the WIMP
and and the nucleon are normalized as
$\psi^\dagger_{\chi,p}\psi_{\chi,p}=M_{\chi,p}$. The SI and SD
WIMP-nucleus cross sections at asymptotically small energies,
$\sigma_{0\chi N}^{SI}$ and $\sigma_{0\chi N}^{SD}$, can be
expressed in terms of the corresponding WIMP-proton cross section,
\beqa \label{eq:x-sections} \sigma_{0\chi
N}^{SI}&=&{M_{red}^2(M_{nuc})\over M_{red}^2(M_p)}
\left[Z + (A-Z) {f_n\over f_p}\right]^2\sigma_{\chi p}^{SI}~, \nonumber \\
\sigma_{0\chi N}^{SD}&=&{M_{red}^2(M_{nuc})\over M_{red}^2(M_p)}
{4(J+1)\over 3J}\left[\langle S_p\rangle + \langle S_n\rangle
{a_n\over a_p}\right]^2 \sigma_{\chi p}^{SD}~. \eeqa
Here, $A$ and $Z$ are the mass number and the charge of the
nucleus with spin $J$, and $f_p$($f_n$) and $a_p$($a_n$) are the
SI and SD WIMP-proton(neutron) couplings, respectively. The
quantities $\langle S_p\rangle$ and $\langle S_n\rangle$ are the
average spins of the proton and the neutron in the nucleus. At
finite momentum transfer one must average the one-nucleon
operators from Eq.~(\ref{eq:chi-lagr}) over the given nucleus
using some nuclear structure model. Including the WIMP velocity
distribution (see Ref. \cite{kam-rev}), one then obtains for the
elastic scattering rate of WIMPs on the nucleus (per unit detector
mass),
\beqa
\label{eq:rate}
{dR\over dE}&=&{\rho_\chi\over 4 v_E M_\chi M_{red}^2(M_{nuc})}
\left[
{\rm erf}\left({v_{min}+v_E\over v_0}\right)-{\rm erf}
\left({v_{min}+v_E\over v_0}\right)
\right]~, \nonumber \\
&\times&\left[ \sigma_{0\chi N}^{SI}F_{SI}^2(E)+\sigma_{0\chi
N}^{SD}{S_A(E)\over S_A(0)}
\right]~, \nonumber \\
v_{min}&=&\sqrt{EM_{nuc}\over 2 M_{red}^2(M_{nuc})}~, \eeqa
where $\rho_\chi$ is the local Galactic halo density \footnote
{We take $\rho_\chi$=0.3 GeV/cm$^3$ \cite{kam-rev}. }, $E$ is
the
recoil energy of the nucleus with mass $M_{nuc}$, and $v_E$ and
$v_0$ are, respectively, velocities of the Earth and the Sun in
the Galactic frame.

The form-factors $F_{SI}(E)$ and $S_A(E)$ in Eq.~(\ref{eq:rate})
depend on the nuclear structure. The SI form-factor can be well
approximated by \cite{vogel92}
\beqa
F_{SI}(E)&=&{3j_1(qR_1)\over qR_1}e^{-(qs)^2/2}~, \nonumber \\
q&=&\sqrt{2 M_{nuc} E}~, \nonumber \\
R_1&=&\sqrt{1.44 A^{2/3}-5s^2}~{\rm fm} ~,\nonumber \\
s&\approx& 1~{\rm fm}~,
\eeqa
where $j_1(x)$ is the spherical Bessel function. Unfortunately, there
is no such universal expression for $S_A(E)$. It can be
parameterized in terms of three nucleus-dependent functions:
\beq
S_A(E)={1\over 4}\left[
(a_p+a_n)^2 S_{00}(E)+(a_p-a_n)^2 S_{11}(E)+(a_p^2-a_n^2) S_{01}(E)
\right]
\eeq
where $S_{ij}(E)$ for most nuclei used in WIMP searches can be
found in Ref. \cite{ressell}.

It is apparent from Eqs.~(\ref{eq:x-sections}) and (\ref{eq:rate})
that the detection rate for non-relativistic WIMPs depends on only
five parameters (in addition to nucleus-dependent constants). We
choose these parameters to be $\sigma_{\chi p}^{SI}$,
$\sigma_{\chi p}^{SD}$, $f_n/f_p$, $a_n/a_p$, and $M_\chi$.
Various extensions of the SM generally reduce the number of
parameters ({\it e.g.}, in the MSSM one normally has
$f_n/f_p\approx 1$ \cite{kam-rev}). In our analysis we do not
impose such restrictions.

\section{Experimental Data}
\label{sec:data}

\subsection{DAMA NaI}

In utilizing the data published by the DAMA collaboration in Ref.
\cite{dama-naI} we adopt the same approach as in Refs.
\cite{kam-vogel,leszek}. In particular, we define the quantity,
\beq
\kappa=\sum_i\left[
{\left(S_{0,i}^{th}-S_{0,i}^{exp}\right)^2\over \left(\Delta
S_{0,i}^{exp}\right)^2}
+{\left(S_{m,i}^{th}-S_{m,i}^{exp}\right)^2\over \left(\Delta
S_{m,i}^{exp}\right)^2}
\right]~,
\eeq
where $S_{0,i}^{th,(exp)}$ and $S_{m,i}^{th,(exp)}$ are,
respectively, the theoretical (experimental) average value and the
annual-modulation amplitude of the detection rate in the $i$th
energy bin. Theoretical predictions for these quantities as
functions of $M_\chi$, $\sigma_{\chi p}^{SI}$, $\sigma_{\chi
p}^{SD}$, $f_n/f_p$, and $a_n/a_p$ can be obtained using Eq.
(\ref{eq:rate}). In this work, we used the experimental values for
$S_{(0,m),i}^{exp}$ given in the first two columns of Table 1 in
Ref. \cite{dama-naI}. The DAMA preferred region, shown in Fig. 4a)
of Ref. \cite{dama-naI}, is well reproduced for
$\kappa\approx$100.

\subsection{DAMA $^{129}$Xe}

The latest results on WIMP searches with a liquid xenon target by
the DAMA collaboration have appeared in Ref. \cite{dama-Xe}. We
take the limits on counts per detector per unit mass per day
(detector rate unit, $dru$) for various energy bins from Fig. 4
and Table I of Ref. \cite{dama-Xe}. In calculating the limits on the
WIMP cross sections we follow the approach of Ref. \cite{smith96}.
We take the central values for the $dru$'s in all energy bins to
be zero and the 90\% CL upper limits on the $dru$'s to be equal to
1.3 times the total error bar on the $dru$'s. We verified that the
90\% CL upper limits appearing in the last column of Table I in
Ref. \cite{dama-Xe} are reproduced in this way with the error bars
from the second column of the same table.

In this approach, the 90\% CL upper limits on the $dru$ in each
energy bin result in an upper limit on the WIMP-nucleon cross
section $\sigma_{\chi N}^{max}(k)$ with the help of
Eq.~(\ref{eq:rate}). The combined upper limit from all energy bins
is obtained using Eq. (15) of Ref. \cite{smith96},
\beq
{1\over (\sigma_{\chi N}^{max})^2}=\sum_k {1\over (\sigma_{\chi
N}^{max}(k))^2}~.
\eeq
We verified that our calculation reproduces the exclusion curves
shown in Figs. 6(a,b) of Ref. \cite{dama-Xe} for both the SI and
the SD cases.

\subsection{EDELWEISS}

We used Ref. \cite{edel} to incorporate the latest results from
the EDELWEISS experiment that used a heat-and-ionization cryogenic
Ge detector. In order to extract the 90\% CL limit on the average
expected number of events $N$ from the null experimental result we
used the Bayesian approach described in Ref. \cite{PDG02}. For our
analysis, we adopt the following prior distribution for $N$,
\beq
\pi(N)=\Biggr\{ {{0\qquad{\rm for}~N<0~,}\atop{1\qquad{\rm for}~N\ge 0~,}}
\eeq
which leads (after normalization) to a PDF for $N$ of the form
$p_0(N)=e^{-N}$. Therefore, the probability for $N$ to be below some value
$N_0$ is:
\beq
P_0(N<N_0)\equiv\int_0^{N_0}p_0(N)dN=1-e^{-N_0}~.
\eeq
Equating the above probability to the confidence level of the EDELWEISS
constraint (90\%), we obtain
$N_0=\ln 10\approx 2.3$.

The predicted number of events in the EDELWEISS detector for fixed
values of WIMP parameters can be related to the detection rate by
integrating Eq.~(\ref{eq:rate}) over the range of energies
accepted by EDELWEISS (20 keV to 64 keV) and multiplying by the
total effective exposure of 11.7 kg-days \cite{edel}. In order
to properly normalize the detection rate we multiply it by a
factor $C_E$. This factor is adjusted such that the following
predictions given in Ref.~\cite{edel} for numbers of events are
reproduced for the pure SI case: for $M_\chi$=44 GeV and
$\sigma_{\chi p}^{SI}=5.4\times 10^{-6}$ pb one has $N=6.2$, and
for $M_\chi$=52 GeV and $\sigma_{\chi p}^{SI}=7.2\times 10^{-6}$
pb one has $N=9.8$. Note that since there is only one coefficient
to determine, the second data point is redundant, and can thus be used
for a consistency check. We find that for $C_E\approx$0.76 both
test points and the EDELWEISS exclusion curve given in Fig. 5 of
Ref.~\cite{edel} are well reproduced. In all the Figures shown below,
the constraints from EDELWEISS data are obtained by setting the
predicted number of events in the detector equal to $N_0=\ln 10$.

\subsection{ZEPLIN-1}

The ZEPLIN-1 experiment (see Ref.~\cite{zeplin-1}) used liquid
Xenon as a medium and relied on pulse-shape discrimination
analysis. Unfortunately, at present there is no publication
available containing detailed experimental results. The absence of
such data prevented us from treating ZEPLIN-1 results in the same
manner as those from the DAMA experiment with liquid $^{129}$Xe
(see above). Instead, we chose an approach similar to the one we
used for the EDELWEISS experiment. We deduced the parameters
necessary for the calculation from Ref.~\cite{zeplin-1}. In
particular, we chose the visible energy to be between 4 keV and 30
keV, with quenching factor $q_{Xe}({\rm Zeplin})=0.2$. These
parameters correspond to recoil energies between 20 keV and 150
keV. In order to calibrate our calculation we introduced an
effective statistics factor, $C_Z$. We found that the SI 90\% CL
ZEPLIN-1 limit shown in Ref. \cite{zeplin-1} is exceptionally well
reproduced for $C_Z\approx$ 8 kg-days.

Due to the absence of published results, the ZEPLIN-1 limit is the most
uncertain input in our calculation. However, the results of our
analysis are robust and are unlikely to change when the full data is
properly included. Indeed, no change in position of the ZEPLIN-1 curve
in Fig. \ref{fig:pure-SD} will move the upper limit on the WIMP
mass above about 25 GeV, which is still in conflict with the
present lower limit on the MSSM lightest-neutralino mass of 37 GeV.
Similarly, even if the ZEPLIN-1 constraint is entirely removed from
the analysis one would still require $f_n/f_p<0$ to achieve
even marginal agreement among all data for the predominantly SI case
(see explanation to Fig. \ref{fig:SI-fine}). As shown below $f_n/f_p<0$ would
be very unusual in the MSSM.

\subsection{Energetic neutrino searches}
\label{sec:indirect}

A promising method for indirect detection of WIMPs is the search
for neutrinos from WIMP annihilation in the Sun and/or the Earth.
Such neutrinos produce upward muons in the Earth via
charged-current interaction. Measurement of (or constraint on)
the flux of
such muons indirectly constrains the annihilation rate of WIMPs,
which is related to the WIMP capture rate by the Sun and/or the
Earth. A detailed review of such indirect WIMP detection with
extensive list of references on the subject can be found in
Ref.~\cite{kam-rev}. In this work, we use the limits on the fluxes
of neutrino-induced upward muons near the surface of the Earth
obtained by Super-Kamiokande \cite{SK1}
\beqa
\Gamma_\mu(\rm sun)\sim \Gamma_\mu(\rm Earth)\lsim10^{-2}~{\rm
m^{-2}~year^{-2}}~.
\eeqa
Limits from other experiments, such as IMB, Baksan, MACRO, AMANDA,
{\it etc.}, are very similar (see Refs.~\cite{kam-rev,kam-vogel}
and references therein). The upward muon flux is related to the
WIMP-proton cross section as
\beqa\label{eq:muon-rates}
\Gamma_\mu^{SI}&=&1.96\times 10^{-13}d\tanh^2\left(t\over
\tau\right)\xi(M_\chi)f^\prime(M_\chi)\left({M_\chi\over {\rm
GeV}}\right)^2\left({\sigma_{\chi p}^{SI}\over 10^{-40}~{\rm
cm^2}}\right)~,\nonumber \\
\Gamma_\mu^{SD}({\rm sun})&=&1.6\times 10^{-2}\tanh^2\left(t\over
\tau\right)\xi(M_\chi)S(M_\chi/M_p)\left({M_\chi\over {\rm
GeV}}\right)\left({\sigma_{\chi p}^{SD}\over 10^{-40}~{\rm
cm^2}}\right)~,\nonumber \\
\eeqa
where $d=3.3\times 10^8~{\rm m^{-2}~year^{-2}}$ for the Sun and
$d=1.7\times 10^8~{\rm m^{-2}~year^{-2}}$ for the Earth, and
$\tau$ is the time scale for equilibration between WIMP capture
and WIMP annihilation. For the Sun, we can take
$\tanh(t/\tau)\approx 1$, whereas for the Earth it is likely
substantially smaller than unity \cite{kam-rev}. The function
$\xi(M)$ is given by Eq.~(9.54), and $S(x)$ by Eq.~(9.21) in
Ref.~\cite{kam-rev}. The function $f^\prime(M)$ is the
generalization of the function $f(M)$ given by Eq.~(9.28) of the
same reference,
\beq
f^\prime(M)=\sum_if_i\phi_i S(M_\chi/M_{nuc}^i)F_i(M_\chi)
{M_{nuc}^{i~3}M_\chi\over M_p^2(M\chi+M_{nuc^i})^2}\left[{Z_i\over
A_i}+\left(1-{Z_i\over A_i}\right){f_n\over f_p}\right]^2~,
\eeq
where the sum runs over the nuclei with mass $M_{nuc}^i$, charge
$Z_i$, and atomic number $A_i$. The quantities $f_i$ and $\phi_i$
are given in Tables 8 and 9 of Ref.~\cite{kam-rev}. The
form-factor suppression $F_i(M_\chi)$ is only important for iron
in the Earth, where it is given by Eq.~(9.23) of
Ref.~\cite{kam-rev}.

For $M_\chi\gg M_p$ the muon flux $\Gamma_\mu^{SD}$ is roughly
independent of $M_\chi$, which leads to the constraint
$\sigma_{\chi p}^{SD}<1.2\times 10^{-4}/\xi$ pb. In the MSSM one
generally has $\xi\ge$0.03 \cite{kam-rev}, and we obtain
$\sigma_{\chi p}^{SD}<4\times 10^{-3}$ pb. The constraints on
$\sigma_{\chi p}^{SI}$ from energetic solar neutrinos and on
$\sigma_{\chi (p,n)}^{SD}$ from terrestrial neutrinos are too weak
to be interesting. On the other hand, the constraint on
$\sigma_{\chi p}^{SI}$ from energetic neutrinos originating in the
Earth is potentially non-trivial. For the case $f_n/f_p\approx
-0.76$ and $M_\chi\approx 50$ GeV (see Section \ref{sec:majorana})
one obtains
\beq \label{eq:SK-constr}\sigma_{\chi p}^{SI}<{3.6\times
10^{-6}\over \xi\tanh^2[t/\tau_\oplus]}~{\rm pb}~. \eeq
As shown below, one needs $\sigma_{\chi p}^{SI}\approx 0.0035$ pb
for this case. In addition, one can deduce from Ref.
\cite{kamion94}
\beq \left({t\over \tau_\oplus}\right)^2=2.4\times
10^{3}f^\prime(50{\rm ~GeV}){\sigma_A v\over 10^{-26}{\rm~cm^3
s^{-1}}} {\sigma_{\chi p}^{SI}\over 1{\rm pb}}~, \eeq
where $\sigma_A$ is the WIMP annihilation cross section in the
limit of zero relative WIMP velocity $v$. In order to have a WIMP
relic density of order unity we must have $\sigma_A v\approx
10^{-26}{\rm~cm^3~s^{-1}}$ \cite{kamion94}, and for
$f_n/f_p=-0.76$ one has $f^\prime(50{\rm ~GeV})=0.85$. Therefore,
$(t/\tau_\oplus)^2\approx 2\times 10^{3}~(\sigma_{\chi
p}^{SI}/1{\rm pb})\approx 8$, or $\tanh^2(t/\tau_\oplus)\approx
1$. Now, Eq.~(\ref{eq:SK-constr}) can be simply rewritten as a
constraint on $\xi$: $\xi<0.001$ for $f_n/f_p=-0.76$. For the
largest value $\xi=0.001$ allowed in this scenario the energetic
neutrino constraint on the WIMP-proton SD cross section is
$\sigma_{\chi p}^{SD}<0.12$ pb.

Can the constraints coming from non-observation of upward muons
from energetic solar neutrinos be evaded? Below we consider two
possibilities. In the first case, the MSSM WIMPs predominantly
decay into light fermions. Because the annihilation rate
$\Gamma(\chi\chi\to f^+f^-)$ is proportional to the fermion mass
squared $m_f^2$, direct annihilation into neutrinos is virtually
impossible, and energetic neutrinos appear in the decay chain of
the initial annihilation products. The total flux of the neutrinos
originating from the branch $\Gamma(\chi\chi\to f^+f^-)$ inherits
the suppression by $m_f^2$, and in the case where only light
fermions appear during annihilation the flux may be orders of
magnitude below the conventional estimates \cite{kam-vogel}.
Effectively, in this case $\xi(M_\chi)$ is substantially smaller
than estimated in Ref.~\cite{kam-rev}, which weakens constraints
on $\sigma_{\chi p}^{SI,SD}$. In the second case, the WIMP is not
identical to the anti-WIMP, and only WIMP--anti-WIMP annihilation
is allowed. Although direct annihilation into neutrinos may be
possible (leading to stronger energetic neutrino signals), the
annihilation rate may still be significantly suppressed in
the presence of significant WIMP-anti-WIMP asymmetry.

\subsubsection{WIMPs only decay into light fermions}
\label{sec:small-xi}

As pointed out in Ref. \cite{kam-vogel} the energetic-neutrino
constraint could be evaded if the WIMPs annihilated to $u\bar u$,
$d\bar d$, $s\bar s$, $e^+ e^-$, and/or $\mu^+\mu^-$ pairs but not
$c\bar c$, $b\bar b$, nor $\tau\bar \tau$ pairs. Such a situation
can in principle be achieved in the MSSM by fine tuning the
sfermion masses and the WIMP composition. Let us write the
lightest neutralino field as
\beq
\chi=Z_1{\tilde B}+Z_2{\tilde W}_3+Z_3{\tilde H}_1+Z_4{\tilde H}_2~,
\eeq
where $Z_i$ are constants subject to $\sum |Z_i|^2=1$, ${\tilde B}$ and
${\tilde W}_3$ are superpartners of the $B$ and $W_3$ gauge bosons, and
${\tilde H}_{1,2}$ are superpartners of the neutral Higgs bosons. Examination
of
the general expressions for the axial-vector and scalar neutralino-fermion
couplings (Eq.~(3.6) in Ref.~\cite{vogel92}) shows that if
$|Z_3|=|Z_4|$ and $Z_2=\tan\theta_W Z_1$ the
scalar coupling vanishes and the axial coupling $a_f$ is inversely proportional
to the fermion superpartner mass squared,
\beq a_f\sim {1\over M_{\tilde f}^2} \eeq
where $M_{\tilde f}$ is the mass of the superpartner of the
fermion $f$. Choosing the superpartner mass to be very large for
the charm and bottom quarks, as well as for the $\tau$ lepton, one
can force the neutralinos to annihilate into the light quarks and
leptons only. However, choosing such flavor non-universal masses
for the scalars may be problematic in the MSSM. Indeed, one must
obey the existing stringent constraints on the size of
flavor-changing neutral currents (FCNC)(see {\it e.g.} Ref.
\cite{fcnc} for a review of FCNC constraints on the MSSM
spectrum). Since evading energetic-neutrino constraints in a way
we just described requires significant flavor non-universality
among certain entries of the squark mass matrices, whereas
smallness of the FCNC prefers the opposite, one may expect that
fine tuning would be required to make any such scenario
phenomenologically viable. We note that even for non-MSSM WIMPs
flavor non-universality of the WIMP couplings is needed to evade
the energetic-neutrino constraint in the described manner, and
experimental limits on FCNC are likely to present complications
for any WIMP candidate.

\subsubsection{The WIMP is not identical to anti-WIMP}

Another way to evade the energetic-neutrino constraint is to
consider WIMPs that are not identical to their antiparticles.
Indeed, for the total number density of the dark matter particles
$N_0$ one has
\beqa
n_0+{\bar n}_0&=&N_0~,\nonumber \\
n_0-{\bar n}_0&=&W~, \eeqa
where $n_0$ and $\bar n_0$ are the WIMP and anti-WIMP densities in
the vicinity of the Solar system, and $W$ is the \lq\lq WIMP
number" density, which may be generated in the presence of some T-, C-
and CP-violating physics. The (anti-)WIMP densities $n$ and $\bar
n$ inside the solar core obey the equations \cite{Griest86},
\beqa \label{eq:wimps-sun}
\dot n&=&An_0-Bn{\bar n}-\Gamma_{esc}~, \nonumber \\
\dot {\bar n}&=&{\bar A}{\bar n}_0-Bn{\bar n}-{\bar\Gamma}_{esc}~,
\eeqa
where the first and the second terms on the RHS of both equations
represent, respectively, the capture rate of WIMPs incident on the
Sun and the annihilation of WIMPs and anti-WIMPs in the solar
core. The constants $A$ and $\bar A$ are proportional,
respectively, to the WIMP-proton and anti-WIMP--proton scattering
cross sections. The last terms represent the losses of WIMPs due
to evaporation (see {\it e.g.} Ref. \cite{Gould87}).

Consider the case where $n_0\gg {\bar n_0}$. In the stationary state $\dot
{\bar n}=0$, and we have for the
annihilation rate $\Gamma_{ann}={\bar A}{\bar n}_0$. We make a
reasonable assumption that annihilation is the dominant mechanism
for anti-WIMP loss if ${\bar n}\ll n$ because the evaporation rate
is suppressed by a small Boltzmann-like factor $e^{-E_e/kT_W}$.
Here, $E_e$ is the escape energy and $T_W$ is the effective WIMP
temperature. On the other hand, for neutralinos in the MSSM we
have ($\Gamma_{esc}$ can be neglected here as well),
\beq \dot N=A^{MSSM}N_0-B^{MSSM}N^2~. \eeq
In the stationary state this would give for the annihilation rate
$\Gamma_{ann}^{MSSM}=A^{MSSM}N_0$. We find that
\beq \Gamma_{ann}={{\bar A}\over A^{MSSM}}{{\bar n}_0\over
N_0}\Gamma_{ann}^{MSSM} ={{\bar A}\over 2
A^{MSSM}}\Gamma_{ann}^{MSSM}\left(1-{W\over N_0}\right) \eeq
We generally expect $A\sim {\bar A}$, and since the WIMP-proton
cross section is constrained by DAMA we expect $A\sim A^{MSSM}$.
Therefore, we conclude that $\Gamma_{ann}\sim
\Gamma_{ann}^{MSSM}\left(1-{W/N_0}\right)$. The flux of upward muons
in the Super-Kamiokande detector is proportional to the WIMP
annihilation rate: $\Gamma_\mu\sim\xi(M_\chi)\Gamma_{ann}$.
Therefore, we obtain,
\beq \Gamma_\mu\sim {\xi(M_\chi)\over \xi^{MSSM}(M_\chi)}
\left(1-{W\over N_0}\right)\Gamma_\mu^{MSSM}~. \eeq
Since the WIMP is not identical to its anti-WIMP, direct WIMP
annihilation into neutrinos may be possible, which
may yield ${\xi(M_\chi)/\xi^{MSSM}(M_\chi)}>1$. On the other hand,
for $1-W/N_0\ll 1$, the muon detection rate in the Sun can still
be significantly below the one predicted by Eq.~(5) in Ref.
\cite{kam-vogel}, and the energetic-neutrino constraint may not
apply. In this work, we first perform the analysis without this
constraint, and then add it later on.

\section{Analysis}
\label{sec:analysis}

\subsection{The energetic-neutrino constraints are not applicable}
\label{sec:dirac}

\begin{figure}[ht]
\begin{center}
\includegraphics[width=5in]{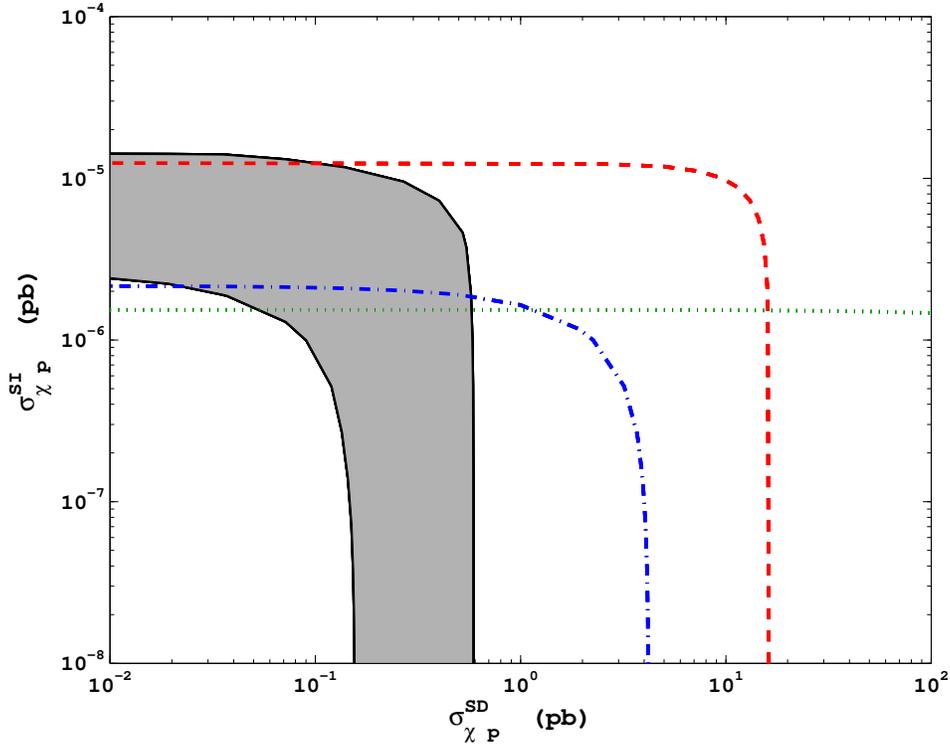}
\caption{Constraints in the $\sigma_{\chi p}^{SI}$ vs. $\sigma_{\chi
p}^{SD}$ plane due to various direct WIMP searches provided that
the energetic-neutrino constraint is not included.  In this
plot, $f_n/f_p=1$, $a_n/a_p=0$, and $M_\chi=50$ GeV.  The shaded
area is the $3\sigma$ allowed region consistent with the annual
modulation observed by the DAMA collaboration in their NaI
detector \cite{dama-naI}.  The region below the dashed curve is the region
allowed by the DAMA $^{129}$Xe experiment
\cite{dama-Xe}.  The region below the
dotted line is that allowed by the EDELWEISS experiment \cite{edel}, and region below the dash-dotted
line is allowed by ZEPLIN-I experiment \cite{zeplin-1}.}
\label{fig:dirac}
\end{center}
\end{figure}

Let us first consider the constraints on the WIMP parameter
space without the energetic-neutrino bounds. In this case an
agreement between all direct search experiments
can be achieved for a predominantly SD WIMP-nucleon interaction for
a wide range of WIMP masses and couplings. A similar conclusion
was originally obtained in Ref.~\cite{kam-vogel}. In Fig.
\ref{fig:dirac} we plot the allowed region in  $\sigma_{\chi
p}^{SI}$ vs. $\sigma_{\chi p}^{SD}$ plane for $f_p/f_n=1$,
$a_n/a_p=0$ and $M_\chi$ =50 GeV. The shaded region is allowed by
all direct search experiments described in Section \ref{sec:data}.
Since a substantial SD interaction is required, the WIMP cannot be a
scalar particle in this case.

\subsection{The energetic-neutrino constraints are applicable}
\label{sec:majorana}

\begin{figure}[ht]
\begin{center}
\includegraphics[width=5in]{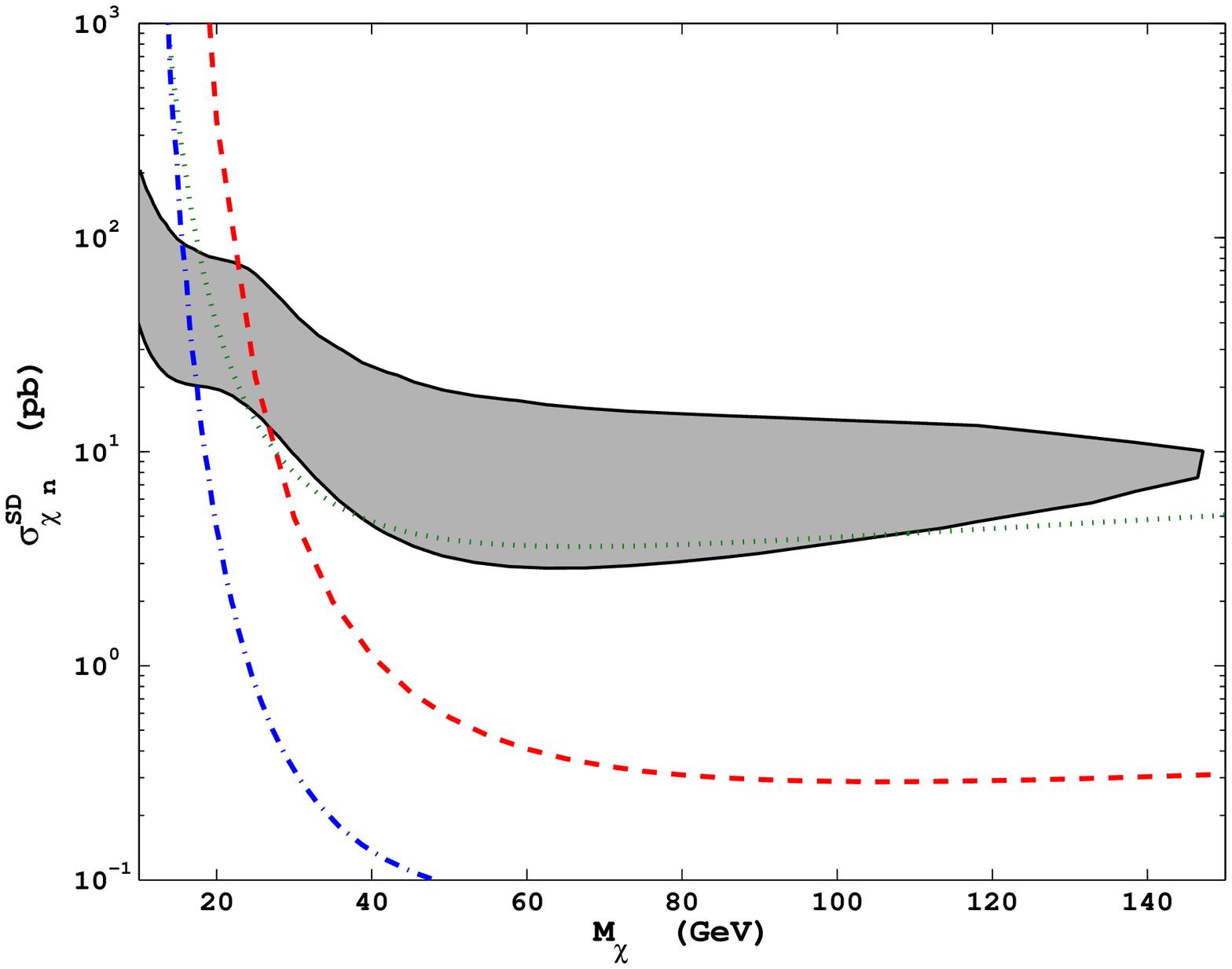}
\caption{Constraints in the $\sigma_{\chi n}^{SD}$(pb) vs.
$M_\chi$ (GeV) plane due to various direct WIMP searches. To
illustrate, we took $a_p/a_n$=0 (pure WIMP-neutron interaction).
This choice automatically ensures that the energetic-neutrino
constraints are satisfied. Notation is the same as in Fig.
\ref{fig:dirac}.} \label{fig:pure-SD}
\end{center}
\end{figure}

We now include the energetic-neutrino constraint. As is well
known, for the usual case $f_n/f_p\approx 1$ the results from all
direct search experiments cannot be reconciled for the pure SI
case at about 3$\sigma$ level, and some SD interaction appears
necessary. With the WIMP-proton SD cross section limited from
above by the energetic-neutrino searches, a relatively large
WIMP-neutron SD cross section is needed to accommodate the
DAMA/NaI result. The situation is similar to the one described in
Ref.~\cite{kam-vogel} for the pure WIMP-neutron interaction, and
the allowed region for this case is shown in Fig.
\ref{fig:pure-SD}. For this figure, we set the WIMP-proton SD
cross section to zero thereby automatically satisfying the
energetic-neutrino constraint.
In this scenario, there is an upper limit on the WIMP mass. The
strongest limit, about 18 GeV, will likely be provided by the
ZEPLIN-1 experiment. We found that this limit does not
significantly change even if some SI interaction is allowed. If
the ZEPLIN-1 result is not included the upper limit on the WIMP
mass increases to about 25 GeV.

It is interesting to note that MSSM neutralinos with mass
below 37 GeV are excluded by direct collider searches
\cite{PDG02}. Although the analysis of such searches is not
completely general since it assumes gaugino and sfermion mass
unification at the GUT scale \cite{PDG02}, evading this limit, if
at all possible, would require fine tuning of the MSSM parameters.

\begin{figure}[ht]
\begin{center}
\includegraphics[width=5in]{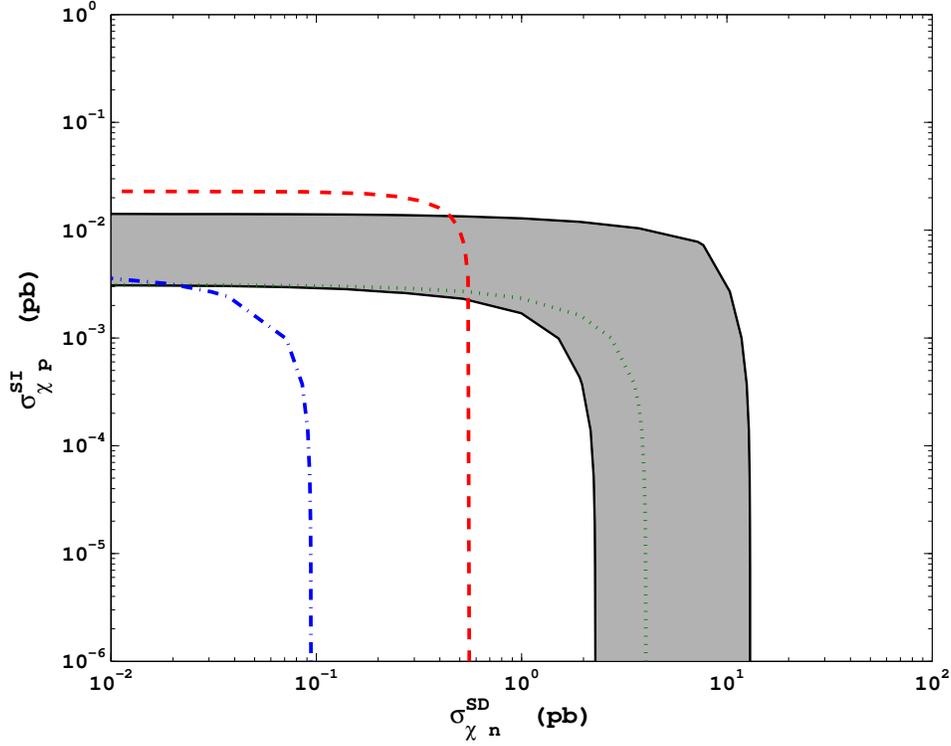}
\caption{Constraints in the $\sigma_{\chi p}^{SI}$(pb) vs.
$\sigma_{\chi n}^{SD}$ (pb) plane due to various direct WIMP
searches. We took $f_n/f_p$=-0.76. Notation is the same as in Fig.
\ref{fig:dirac}. The energetic-neutrino constraint for this case,
$\sigma_{\chi p}^{SD}<0.12$ pb (see Section \ref{sec:indirect}),
is satisfied.} \label{fig:SI-fine}
\end{center}
\end{figure}

Up to this point, we always maintained the condition $f_n/f_p=1$.
It turns out that relaxing this constraint allows one to achieve
marginal agreement among all data for $M_\chi$ in the vicinity of
50 GeV. However, with the ZEPLIN-1 result included this occurs for
only a very narrow range of $f_n/f_p$. The possibility arises
because the neutron-to-proton ratio differs slightly from one
nucleus to the other.  Specifically, if $-0.77\lsim f_n/f_p\lsim
-0.75$, then the WIMP coupling to the neutron in Xe cancels that
from the proton, allowing a null result in ZEPLIN-1 to be
consistent with a modulation in DAMA.  If ZEPLIN-1 is not
considered the allowed range is increased but the consistency
range for $f_n/f_p$ is still very narrow. The situation for
$f_n/f_p=-0.76$ is shown in Fig. \ref{fig:SI-fine}. The small
region allowed by all the data is centered around $\sigma_{\chi
p}^{SI}\approx$0.0035 pb. Note that, due to significant
cancellation between the WIMP-neutron and the WIMP-proton
scattering amplitudes, the individual WIMP-nucleon cross sections
are about $10^2$ times larger than in the case where $f_n/f_p=1$.
The inverse of this number, 0.01, reflects the amount of fine
tuning required for this solution to work.

{}From the standpoint of the MSSM, this solution is in fact \lq\lq doubly
fine-tuned". In addition to making sure that $f_n/f_p$ is
carefully selected to fit all the data one has to tune the model of
SUSY breaking to even obtain $f_n/f_p<0$ (normally, one has $f_n\approx f_p$ in
the MSSM \cite{kam-rev}). It is interesting to ask, therefore, whether
a significant deviation from the approximate equality
$f_n/f_p\approx 1$ is at all possible within the MSSM. In general,
we have,
\beq f_N=f_0+f_1{\hat \tau}_3~, \eeq
where $N=p,n$ and ${\hat \tau}_3$ is the usual Pauli matrix in the
strong isospin space. The Feynman diagrams and detailed
expressions for $f_{p,n}$ can be found in Ref. \cite{kam-rev}.
Since only the up and down quarks have non-zero strong isospin,
the isovector part of $f_N$ is entirely due to coupling of the
neutralino bilinear to isovector operators built from the up and
down quarks. The isoscalar part arises due to coupling to various
isoscalar operators built from the up and down quarks, the strange
and all heavy quarks, and the gluons. The scalar coupling of the
neutralino to a quark has the form \cite{kam-rev}\footnote{We
assume that the squark mass matrices are diagonal in the flavor
space.},
\beq {\cal L}_{\chi,q}=G_F {m_q \over M_W}{\bar \chi}\chi ~{\bar
q}q\left(A_q{M_W^2\over M_h^2}+B_q{M_W^2\over {M_{\tilde q}^2
-(M_\chi+m_q)^2}}\right)+{\cal O}({1\over M_{\tilde q}^4}) \approx
m_q f_q{\bar \chi}\chi ~{\bar q}q~, \eeq
where $G_F$ is the Fermi constant, $A_q$ and $B_q$ are
dimensionless constants, $m_q$ is the quark mass, $M_{\tilde q}$
is the mass of the superpartner of $q$, $M_h$ is the lightest
Higgs-boson mass, and $M_\chi$ is the neutralino mass. We can now
write the couplings $f_{p,n}$ in the form \cite{kam-rev},
\beqa f_{p,n}=m_{p,n}\left[f_{Tu}^{p,n}f_u+f_{Td}^{p,n}f_d
+f_{Ts}^{p,n}f_s+\cdots\right], \eeqa
where \lq\lq$\cdots$" stand for both the isovector terms of order
$1/M_{\tilde q}^4$ and all remaining isoscalar contributions.
Here, $m_Nf_{Tq}^{N}=\langle N|m_q{\bar q}q|N \rangle$. Ignoring
the remaining terms we obtain (see Table 6 in Ref.
\cite{kam-rev}),
\beq \left|{f_1\over f_0}\right| \lsim \left|
{f_u(f_{Tu}^p-f_{Tu}^n)+f_d(f_{Td}^p-f_{Td}^n) \over 2f_{Ts}f_s
+f_u(f_{Tu}^p+f_{Tu}^n) +f_d(f_{Td}^p+f_{Td}^n)} \right|
\approx\left|{-0.014f_u+0.025f_d\over f_s+0.15 f_u+0.27
f_d}\right|~. \eeq
In most models of SUSY breaking $f_{u,d}\sim f_s$, and although
the above range is nothing more than an estimate it at least shows
that $|f_1/f_0|\ge 1$ (which would lead to a negative ratio
$f_n/f_p$) is generically disfavored in the MSSM. However, it is
possible in principle to fine-tune the parameters to obtain
$|f_1/f_0|\ge1$. One scenario arises when $|f_s|\ll |f_{u,d}|$.
This can be achieved if $M_{\tilde q}$ is very large for all
squarks except for $\tilde u$ and $\tilde d$, all physical Higgs
bosons, except the lightest one, are very heavy, and coupling
between the WIMP and the lightest Higgs boson is tuned to zero. In
this case, one finds (we use the MSSM Feynman rules given in
Ref.~\cite{rosiek} and bring the notation in correspondence with
Ref.~\cite{vogel92}),
\beqa \label{eq:fuds} f_u&=&-{g^2\over 4M_W M_{\tilde
u}^2\sin\beta}Z_4\left(Z_2-\tan\theta_W Z_1\right)~,
\nonumber \\
f_d&=&{g^2\over 4M_W M_{\tilde d}^2\cos\beta}Z_3\left(Z_2-\tan\theta_W
Z_1\right)~,
\nonumber \\
f_s&\approx&0~.
\eeqa
Here, $g$ is the $SU(2)_L$ gauge coupling, and $\tan\beta$ is the ratio of the
expectation values for the two Higgs boson doublets in the MSSM. We find
\beq
\left|{f_1\over f_0}\right|\sim \left|{0.014+0.25 \tan^2\beta{M_{\tilde
u}^2\over M_{\tilde d}^2}\over -0.15+0.27\tan^2\beta{M_{\tilde u}^2\over
M_{\tilde d}^2}}\right|~.
\eeq
If ${M_{\tilde d}/M_{\tilde u}}\approx 1.3\tan\beta$ then one can
have $|f_1/f_0|>1$, which may lead to $f_n/f_p<0$. At present,
$\tan\beta\gsim 3$ is favored by precision data \cite{PDG02}.
Therefore, one needs mild hierarchy between the up- and
down-squark masses to obtain $f_n/f_p<0$: ${M_{\tilde d}\gsim
3.9M_{\tilde u}}$.

There might be other scenarios yielding $f_n$ substantially
different from $f_p$. However, they will all have to share the
same property: the first generation of squarks must be singled out
from the rest to enhance the up- and down-quark contributions to
$f_{0,1}$. Reconciling such significant flavor non-universality of
the squark flavors with stringent constraints from FCNC will
generally require fine-tuning. In this sense, having significantly
different values for $f_n$ and $f_p$ is unnatural, although
potentially possible, in the MSSM. There is an additional
complication, however. For this scenario to work, the WIMP-proton
scattering cross section must be roughly 100 times larger than for
the case $f_n\approx f_p$. On the other hand, one must have
$\xi<0.001$ in this case (requiring $Z_2-\tan\theta_W Z_1\ll 1$,
see Section \ref{sec:small-xi}), which suppresses $f_{u,d}$
according to Eq.~(\ref{eq:fuds}). Partly due to this suppression,
we have found no scenario where sufficient enhancement of
$f_{p,n}$ would occur in the MSSM for $M_\chi\approx 50$ GeV
together with $f_n/f_p\approx -0.76$.

We point out that $f_n/f_p\approx -0.76$ does not hold for any of
the dark-matter candidates we considered. For scalar neutrinos and
KK neutrinos one finds (see Subsection~\ref{sec:examples})
$f_n/f_p=T_n^3/(T_p^3-2Q_p\sin^2\theta_W)\approx -10$. A detailed
study of this ratio for KK excitation of the hypercharge gauge
boson shows that it is at most few percent away from unity. In
summary, although there exists a small phenomenologically allowed
region of parameter space where predominantly SI WIMP-nucleon
interactions can marginally account for all data available on WIMP
searches this region appears to be out of reach for all of the
WIMP candidates we considered.

\section{Conclusions}
\label{sec:conclusions}

In this work we performed a generalized analysis of the
dark-matter detection experiments listed in Section
\ref{sec:data}. Our analysis is formulated in terms of the WIMP
mass $M_\chi$, the SI and SD WIMP-proton cross sections
$\sigma_{\chi p}^{SI,SD}$, and coupling ratios $f_n/f_p$ and
$a_n/a_p$. We found several regions in this parameter space that
allow for agreement among all data.
\begin{itemize}
\item If the energetic-neutrino constraint does not apply it is
possible to reconcile all data with a predominantly SD WIMP-proton
interaction. This scenario can in principle be tested in a direct
dark matter detection experiment with an odd-proton target, such
as {\it e.g.} $^{19}$F \cite{ellis-F19}. Evading the
energetic-neutrino constraint is problematic in the MSSM because
it generally requires highly flavor non-universal squark masses.
Another possibility is a non-self-charge-conjugate WIMP, which
would indicate physics other than the MSSM. In order for this
possibility to be realistic, such physics must accommodate
WIMP-number--violating operators and possess a sufficient amount
of C and CP violation to generate a fractional WIMP asymmetry near
unity.
\item If the energetic-neutrino constraint applies then the SD
WIMP-proton cross section is constrained to be $\sigma_{\chi
p}^{SD}<1.2\times 10^{-4}/\xi$ pb, and for a wide range of
$f_n/f_p$, a SD WIMP-neutron cross section $\sigma_{\chi
n}^{SD}\gsim 30$ pb is required. In addition, $M_\chi$ is
constrained to be below 18 GeV, which is inconsistent with the
present lower limit on the lightest neutralino mass within most
models of SUSY breaking \cite{PDG02}.
\item A heavier WIMP (about 50 GeV) is allowed for a narrow range
$-0.77\lsim f_n/f_p\lsim -0.75$ with essentially no SD interaction
present. In addition, one must have $\sigma_{\chi p}^{SI}\approx
0.0035$~pb. None of the WIMP candidates we considered could
satisfy both requirements at the same time.
\end{itemize}
We see that although it is possible to reconcile the direct and
indirect dark-matter search experiments listed in Section
\ref{sec:data} the resulting parameters are in general unnatural
for the MSSM with most models of SUSY breaking that appear in
the literature. Either
significant flavor non-universality among the SUSY breaking
parameters ({\it e.g.} to evade the energetic-neutrino constraint)
or some delicate relationships among them (to avoid the lower
limit on the lightest neutralino mass) appear necessary.
Generating the WIMP parameters in any of the allowed regions may
be problematic in a theory with non-MSSM WIMPs as well. For
instance, evading the energetic-neutrino constraint in a theory
with non-self-charge-conjugate WIMPs may require a significant
amount of CP-violation to generate a fractional WIMP number close to
one. Such CP-violating interactions may be strongly constrained by
existing limits on {\it e.g.} neutron and atomic electric dipole
moments \cite{PDG02}. If the energetic-neutrino constraint is
adopted, one would have to explain how very light WIMPs
($M_\chi\lsim 18$ GeV) with significant WIMP-quark couplings have
so far escaped detection.

On the other hand, if the DAMA result is removed from the analysis
the MSSM neutralino (as well as other WIMP candidates) will be
compatible with the remaining experiments and still be capable
of being produced with sufficient abundance to account for all
dark matter \cite{PDG02}. In view of this situation, one can
hope that forthcoming direct dark-matter search experiments,
potentially capable of  improving the present sensitivity by
several orders of magnitude, will be able to conclusively confirm
or exclude the results published by DAMA based on their
observation of annual modulation of the detection rate.

\begin{acknowledgments}
We thank Petr Vogel and Gary Pr\'{e}zeau for useful comments. This
work was supported in part by NASA NAG5-9821 and DoE
DE-FG03-92-ER40701 and DE-FG03-02ER41215.
\end{acknowledgments}

\end{document}